\documentclass[a4paper,11pt]{article}
\pdfoutput=1
\usepackage[english]{babel}
\usepackage{hyperref} 
\usepackage[pdftex]{graphicx}

\begin{document}

\pagenumbering{arabic}
\titlepage {
\title{Time over threshold in the presence of noise}
\author{ F. Gonnella$^1$, V. Kozhuharov$^{1,2}$,
 M. Raggi$^1$ \\ 
\\
{\it $^1$INFN Laboratori Nazionali di Frascati, 00044 Frascati (Rome), Italy } \\
{\it $^2$University of Sofia ``St. Kl. Ohridski'', 1164 Sofia, Bulgaria}\\
}
\date{}
\maketitle

\abstract{The time over threshold is a widely used quantity to describe signals
from various detectors in particle physics. Its electronics implementation 
is straightforward and in this paper we present the studies of its behavior 
in the presence of noise. A unique comb-like structure was identified in the 
data for a first time and was explained and modeled successfully. 
The effects of that structure on the efficiency and resolution are also discussed.
}


}

\section{Introduction}



\begin{figure}[!htb]
  \begin{minipage}[t]{0.3\textwidth}
    \centering\includegraphics[width=\textwidth]{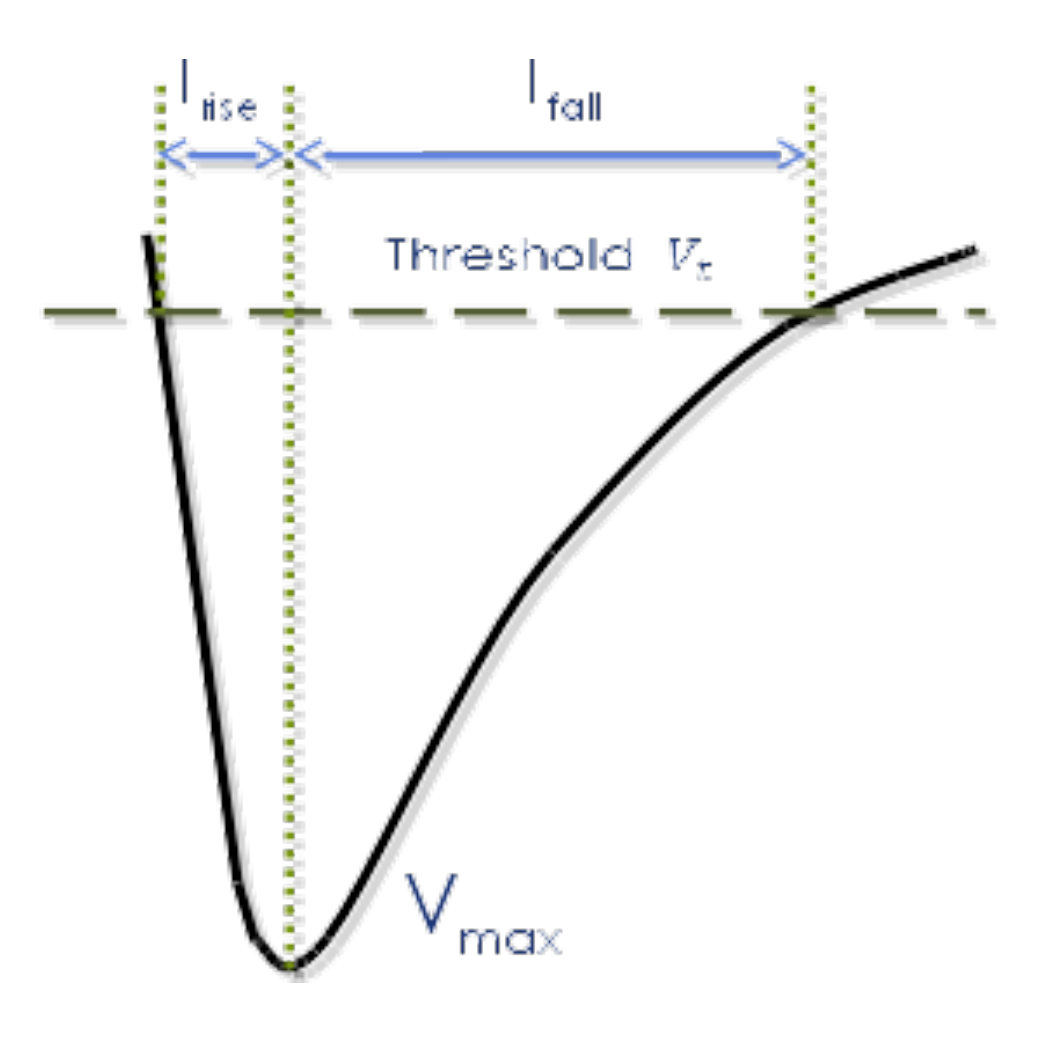}
    \caption{\it Time over threshold }
    \label{fig:tot}
  \end{minipage}\hfill
  \begin{minipage}[t]{0.53\textwidth}
    \centering\includegraphics[width=\textwidth]{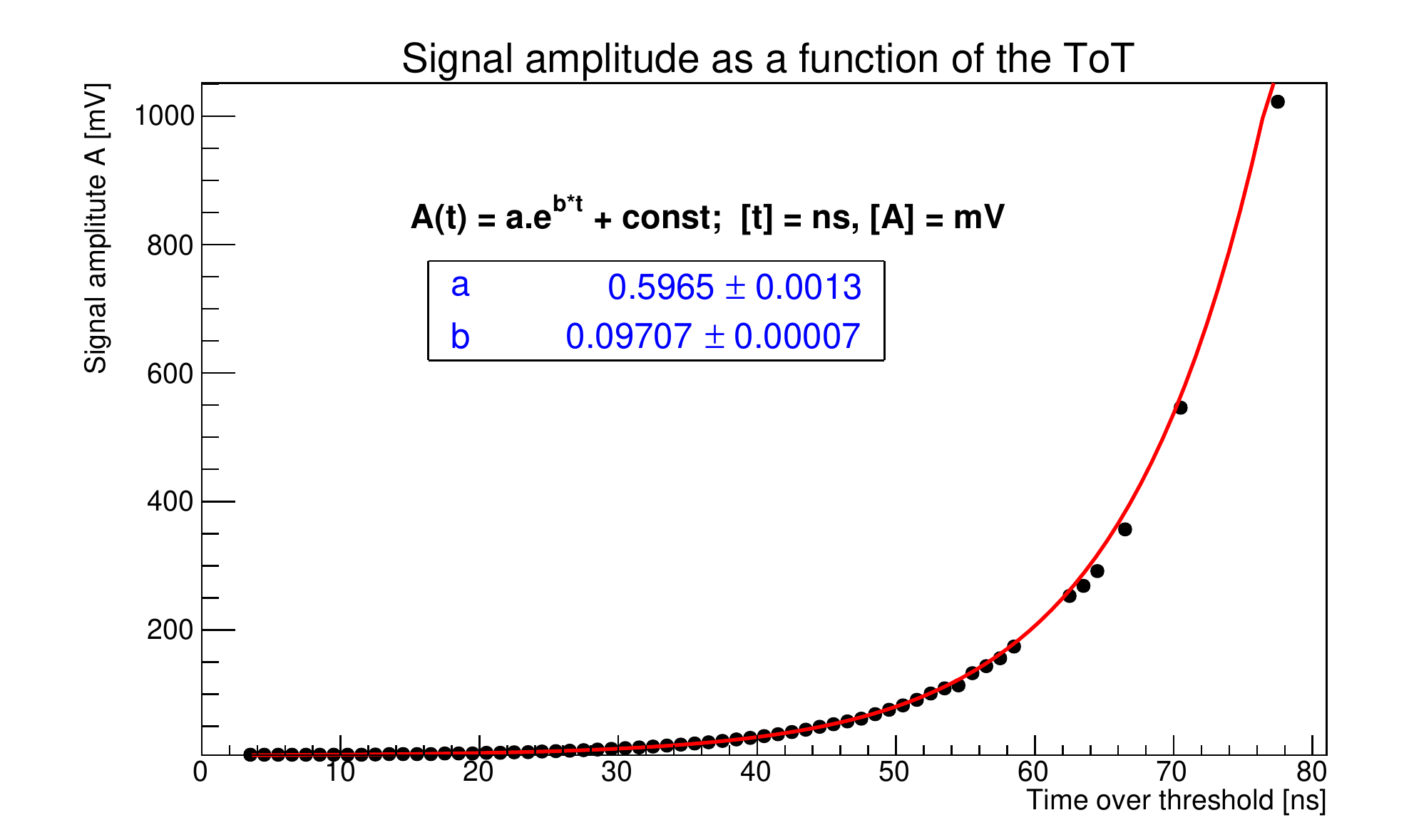}
    \caption{\it Relation between the time over threshold and signal amplitude. The relation
between the signal amplitude and the deposited energy is linear.}
    \label{fig:tot-amp}
  \end{minipage}
\end{figure}

A common setup in particle physics is to couple a particle-sensitive material (e.g. Scintillator, 
Cerenkov) to a photosensitive device (photomultiplier, multipixel photon counter) and measure the
energy deposited in the material by reconstructing a given property of the output pulse - the total
charge collected, the pulse amplitude, etc. 
The measurement of the time over threshold (ToT), as shown in Fig. \ref{fig:tot},    
is composed of two measurements of time for the signal going above (leading) and returning below (trailing) 
a given threshold. 
This provides information about energy deposited by the interacting particle through the 
reconstruction of the difference between leading and 
trailing time $\Theta = t_{trail} - t_{lead}$.
In addition the impact time could also be obtained from the leading time 
with a possible energy dependent correction. 
The dependence of the deposited energy on ToT (see Fig. \ref{fig:tot-amp}) has an exponential form and 
could be parametrized by
\begin{equation}
 E(\Theta) = \alpha~ Q (\Theta) = \alpha~\beta ~A(\Theta) = k * (a.e^{b*\Theta} + const),
\end{equation}
due to the linear relation between energy, charge and signal amplitude
 ($\alpha$, $\beta$, and $k$ are constants). 

The advantage of using the time over threshold instead of charge or amplitude measurement is
the wider dynamic range accessible due to the logarithmic dependence on the energy.
In addition the measurement of the time is performed using time to digital converters (TDCs)
which provide less expensive
solution per channel than the analog to digital converters, 
especially where high signal rate and short signals are expected.

While the charge or amplitude measurements is a well established and mature technique 
the ToT measurement is just becoming attractive nowadays due to the development of
high precision time measurement devices - tens of picoseconds.

In the present article we describe the observation of a peculiar structure in the 
reconstructed distribution of the ToT which we could only explain by the superposition 
of a small amplitude sinusoidal noise on top of the PMT signal.

The charge measurement done by the QDC is completely immune to this effect as the integral 
of a sinusoidal function is zero.




\section{Experimental setup}

The present study was done at LNF-INFN as part of the development of the readout system 
of the Large Angle Photon vetoes for the NA62 experiment at CERN SPS \cite{bib:na62tdr}. 

The NA62 experiment aims to perform a 10\% measurement of the branching fraction 
of the extremely rare decay $K^+ \to \pi^+ \bar{\nu}\nu$ and subsequently to measure 
the $V_{td}$ CKM matrix element. The theoretical prediction for that value is
$Br(K^+ \to \pi^+ \bar{\nu}\nu) =(7.81\pm0.80)*10^{-11}$. The very low rate 
requires an efficient veto of all the other charged kaon decay modes most of which 
contain photons in the final state.
The usage of a 75 GeV kaon beam increases the minimal photon energy at 
which high rejection factor is necessary but still an inefficiency less that 
$10^{-4}$ is required for photons with energy of 100 MeV.
\begin{figure}[htb]
\centering
\includegraphics[width=14cm]{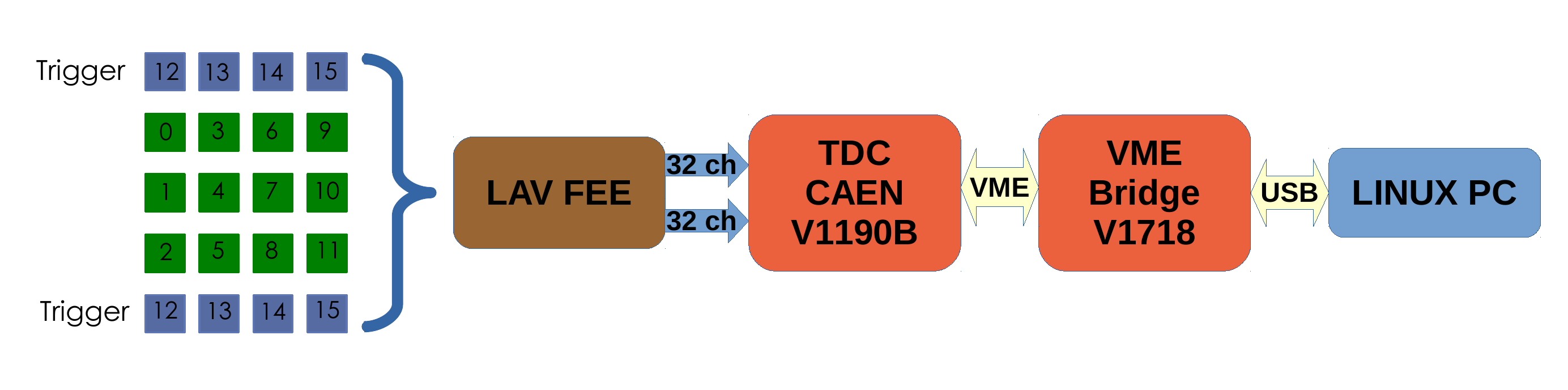}
\caption{Schematics of the experimental setup for the measurement of the ToT distribution
for cosmic rays with Lead-glass block. }
\label{fig:exp}
\end{figure}

The Large Angle Photon vetoes \cite{LAV} consist of lead glass blocks made of 
Schott SF57 lead glass and are coupled to a R2238 photomultiplier. The blocks are arranged 
in the form of rings. A total of 12 Large Angle Photon vetoes were produced, 
with 5 or 4 rings for a total of about 2500 analog channels. The signal from a 100 MeV photon
in the lead glass after propagation through the cables to the front end electronics
could have an average amplitude as low as 10 mV and is almost equivalent to 
the response to the energy deposited by a minimally ionizing particle passing (MIP) through the 
crystal. 

The front end electronics of the lead glass blocks 
was developed
at the LNF \cite{bib:lav-fee}. It is based on a 9U VME mother board receiving 32 
analog inputs. Each signal is clamped, amplified and split into two 
before being transferred to a high speed comparator with an LVDS output driver. 
The comparator threshold could be set through a board controller mezzanine, which provides 
a serial USB and a CAN-Open communication. 
The minimal effective threshold for all the channels was found to be less than 5 mV. 
An additional negative feedback circuit was implemented to dynamically decrease 
the absolute value of the threshold just after the leading edge of the signal.
Such a mechanism, referred to as hysteresis, provides a safety margin against
fast changing signals which would cause the digital LVDS output to oscillate.

The experimental setup used during these studies is shown in Fig. \ref{fig:exp}.
The signals coming from the lead-glass blocks, after the discrimination, 
were readout by a V1190B TDC module which is
based on the CERN HPTDC chip \cite{bib:HPTDC} and incorporates 2 times 32 input channels. 
The data is transferred to a PC for further analysis
 through a V1718 VME controller via a USB connection. 

The recorded data showed a peculiar shape of the ToT distribution. An explanation 
based on the addition of a sinusoidal noise was employed and verified by means
of a numerical signal simulation.

\section{Signal and noise modeling} \label{sec:sig-model}


The general function describing the output signal would be

\begin{equation}
 A(t) = \int_{0}^{t} I(t-\theta) * f(\theta) ~ d\theta,
\end{equation}
where $I(t)$ is the intensity of the light produced in the active material 
and $f(t)$ is the photodetector response function to single electron. The form of the 
light intensity was chosen as an exponential decay with decay time $\tau$, 
\begin{equation}
 I(t) = \frac{N_{0}}{\tau} e^{-t/\tau},
\label{eq:scint}
\end{equation}
assuming that the energy inside the 
active media is released instantly (true for small sized detectors) and the 
only contribution comes from light propagation or scintillating centers decay. 
The normalization $N_{0}$ is the number of the total photons emitted. 

\begin{figure}[!htb]
  \begin{minipage}[t]{0.30\textwidth}
    \centering\includegraphics[width=\textwidth]{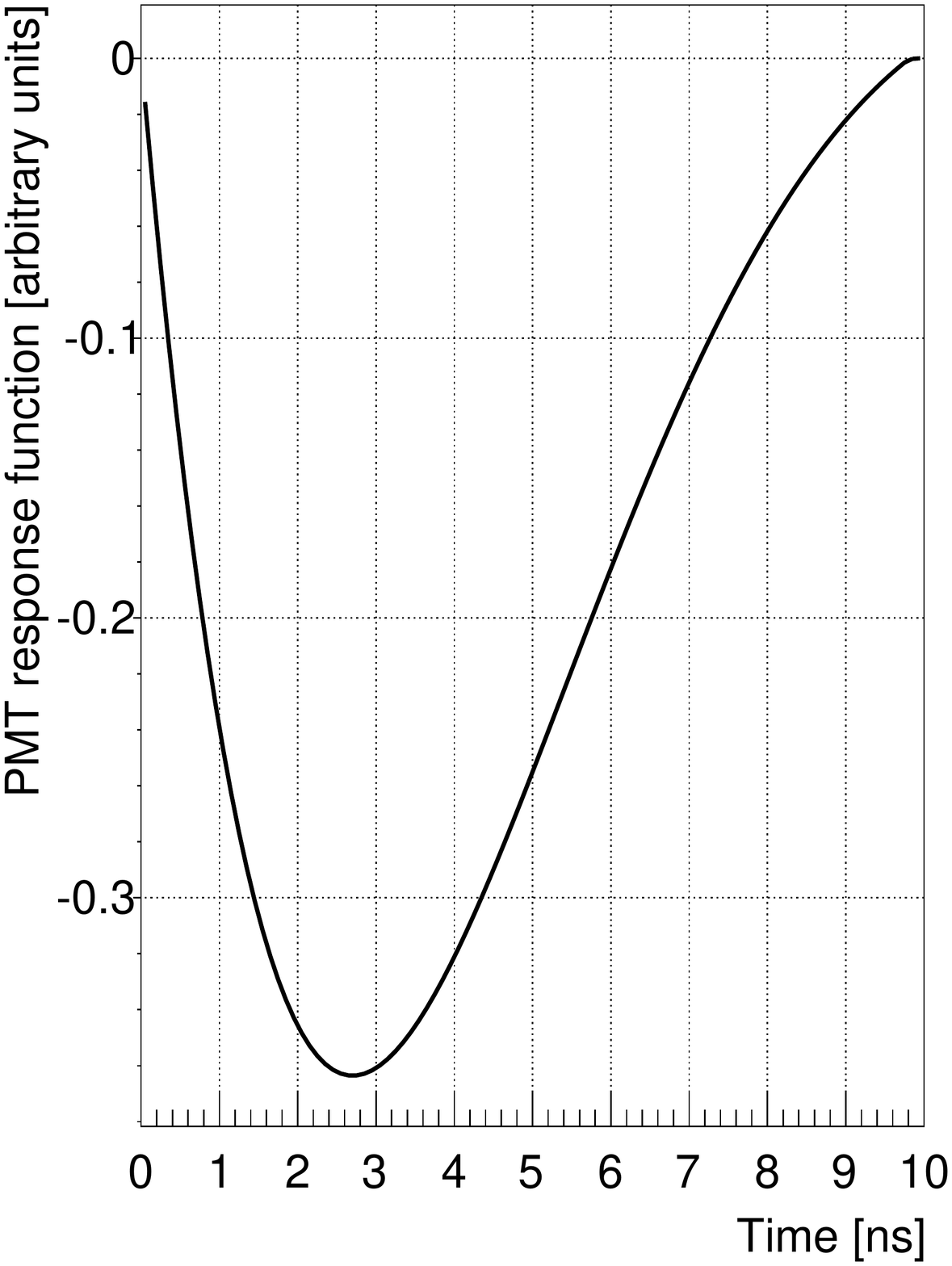}
    \caption{\it Single electron response function of the PMT }
    \label{fig:pmt-ser}
  \end{minipage}\hfill
  \begin{minipage}[t]{0.66\textwidth}
    \centering\includegraphics[width=\textwidth]{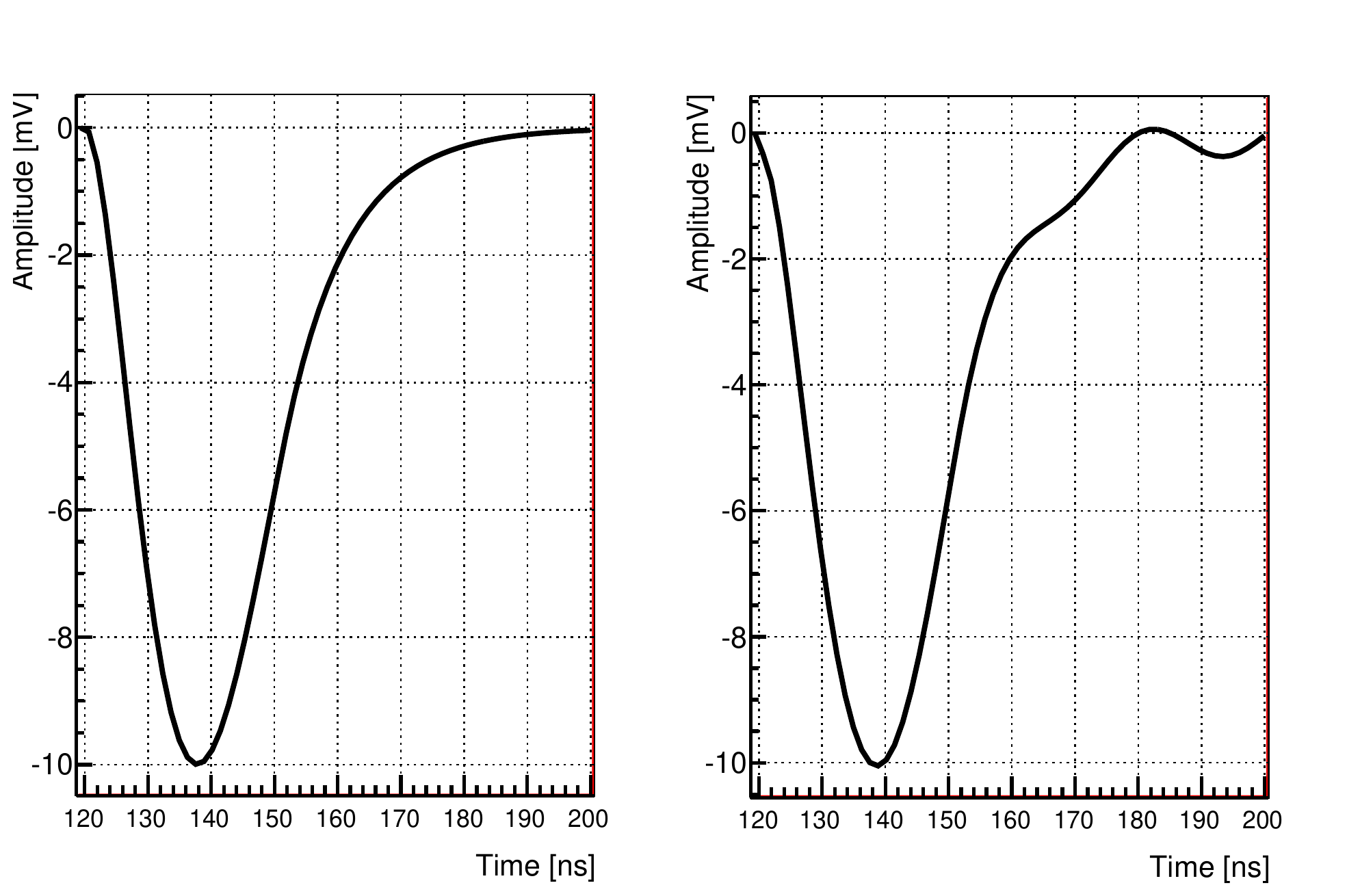}
    \caption{\it Shape of the output signal of the system Lead-Glass block - PMT 
without and with the addition of 300 $\mu$V noise.}
    \label{fig:sim-lav-signal}
  \end{minipage}
\end{figure}

The PMT single electron response was approximated with the function
\begin{equation}
f(\theta) = sin(a\theta)\times e^{-\theta/b},
\label{eq:pmt-ser}
\end{equation}
where $a$ and $b$ were taken as free parameters. 
The use of such a function could be justified with an initial increase of the signal 
amplitude due to arrival of first electrons and further a decrease of the amplitude
due to full charge collection at the anode of the PMT. 

An advantage of using functions \ref{eq:scint}  and \ref{eq:pmt-ser} is 
the simple and analytic form of the final signal. 
The resulting output signal amplitude at the anode of the PMT would then be described as 
\begin{equation}
 A(t) = \frac{-N_{0}}{a^2 + c^2} \times \frac{e^{-t/\tau}}{\tau}\times
\left[e^{ct}(c~sin(at) - a~cos(at)) + a \right], 
\end{equation}
where $c = 1/\tau - 1/b$. The width of the signal $\Delta T$ is described by the parameters
$a$ - $\Delta T = \pi/a$. 

This signal model had been previously 
applied 
to describe the Eljen 212 scintillator coupled
to Hamamatsu R6427 photomultiplier. The scintillator decay time constant, 
the PMT rise time and fall time were found to be consistent with the specification. Their behavior
with the different PMT voltages were as expected. 
This check lead to the confidence of applying the chosen signal description 
to model the time over threshold behavior in various conditions.

The noise was simulated by adding a parasitic signal  
\begin{equation}
 A_{tot}(t) = A(t) ~+ ~A_{noise}(t) = A(t)~ + ~A_{0}~ sin(2\pi f t + \phi),
\end{equation}
where $A_{0}$ is the noise amplitude, $f$ is the noise frequency and $\phi$ is a
random chosen phase. The noise could either be picked-up from external sources
or generated internally in the front end electronics by parasitic positive 
feedback. For the present studies the real origin of the noise is not important.

Thus the basic parameters used to describe the characteristics of the output signal were:

\begin{itemize}
 \item 
{\bf Average signal amplitude.} The signal amplitude was simulated as a Landau distribution
with most probable value of 10 mV and a gaussian sigma of $\approx$2 mV. This corresponds to a
total of 25 photoelectrons per MIP from the Lead-glass - PMT system.

 \item 
{\bf Threshold. } The threshold was kept fixed during the simulation in order to see what was the 
additional effect of the hysteresis and the noise on top of the PMT signal. 
Two values were studied as examples - 5 mV and 7 mV. 

 \item 
{\bf Hysteresis. } The hysteresis is an important ingredient of the time over threshold 
circuit and it prevents short and oscillating output when the input signal 
is very close to the threshold. The hysteresis was varied from 0 mV to 3 mV in steps of 300 $\mu$V.

 \item 
{\bf Noise amplitude.} The noise amplitude was varied from 0 mV to 3 mV in steps of 300 $\mu$V. 

 \item 
{\bf Noise frequency.} The noise frequency was kept fixed to 300 MHz as was independently observed with a digital oscilloscope.
\end{itemize}

The output quantity is the time over a certain threshold. 
A signal is considered to be detected if the time over threshold 
is longer than a fixed minimal time $\Theta_{0}$. In the present 
studies $\Theta_{0} = 5$ ns was used, since it was compatible with the dead 
time of the HPTDC.

\section{Results and discussions}

\begin{figure}[!htb]
  \begin{minipage}[t]{0.4\textwidth}
    \centering\includegraphics[width=\textwidth]{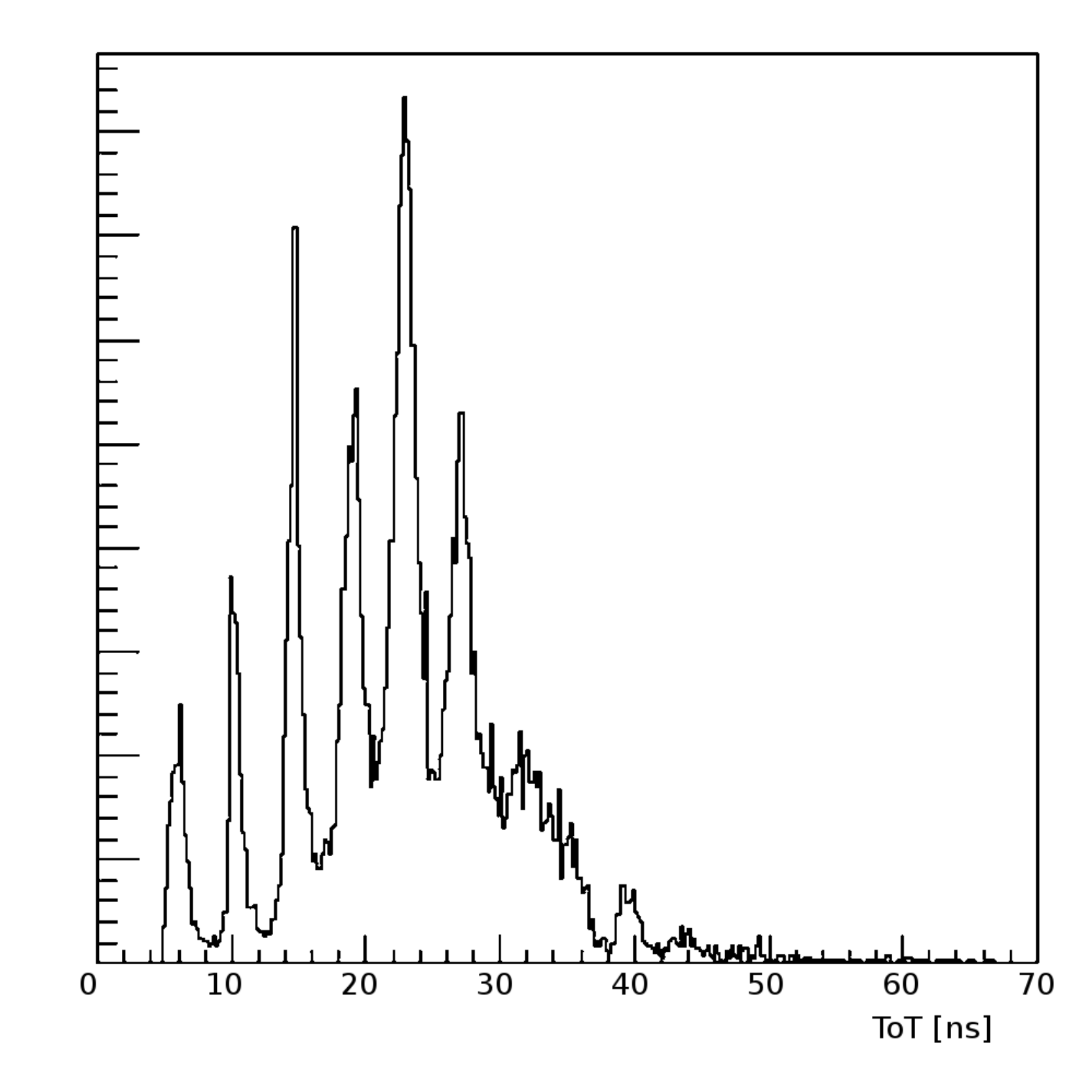}
    \caption{\it Measured time over a 10 mV threshold distribution for cosmic rays from a Lead-Glass block.  }
    \label{fig:data-tot}
  \end{minipage}\hfill
  \begin{minipage}[t]{0.55\textwidth} 
    \centering\includegraphics[width=\textwidth]{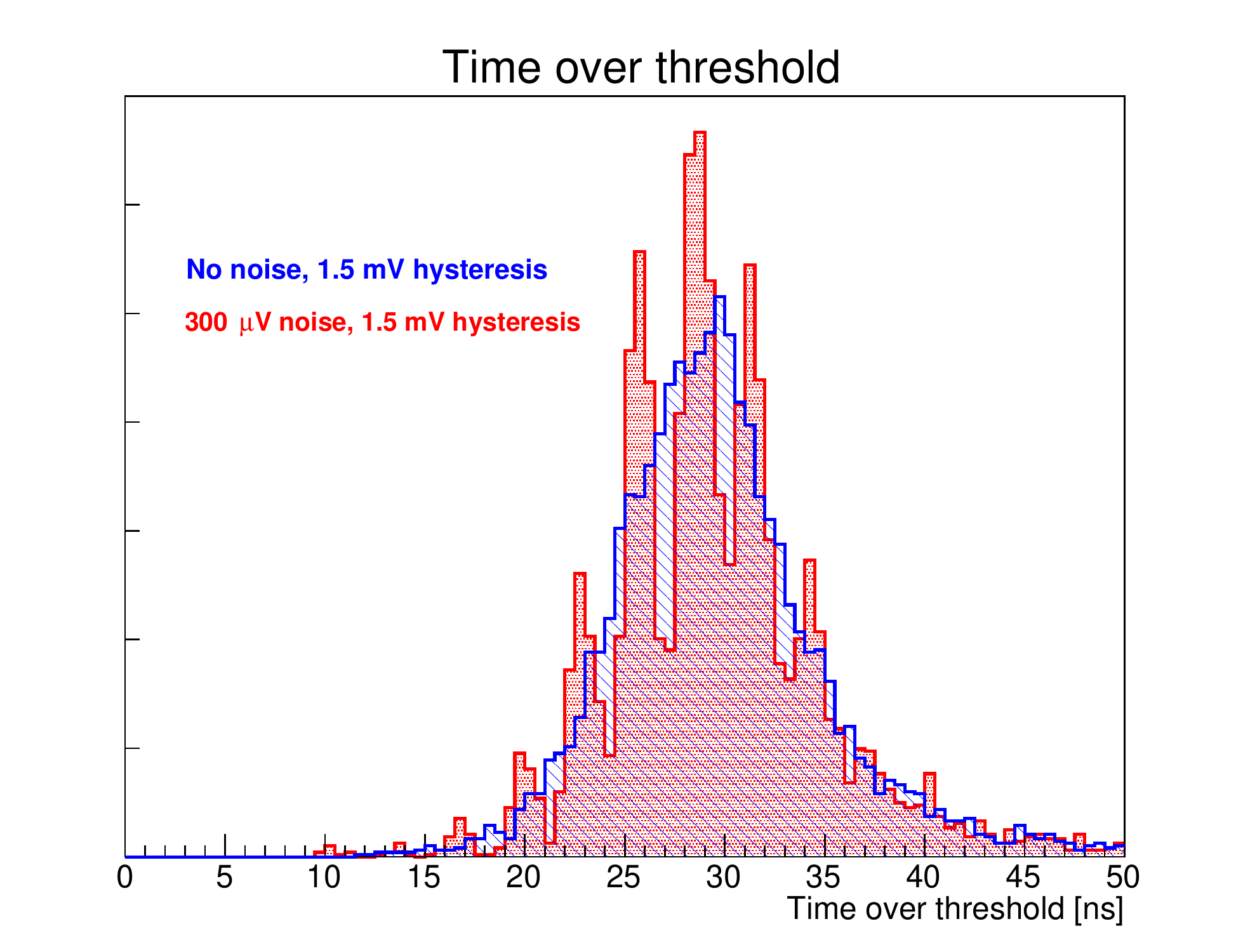}
   \caption{\it Simulated time over 5 mV threshold distribution for cosmic rays with 
  external noise with amplitude of 300 $\mu$V
}
    \label{fig:sim-real-data}
  \end{minipage}
\end{figure}

With the described experimental setup the obtained time over threshold 
distribution is shown in Fig. \ref{fig:data-tot}. It possesses a 
bizarre feature of multiple peaks - a comb-like structure - 
which was initially puzzling and stimulated the presented study.
Few explanations were considered ranging from effects due to energy deposit 
and photoelectron emission to TDC miss-functioning effects (stuck bit for example). 
Finally the data was reproduced exploiting the signal modeling described in section \ref{sec:sig-model}
with 300 MHz sinusoidal external (pick-up) noise on the input analog signals.

The effect on the inclusion of the extra noise is shown in Fig. \ref{fig:sim-real-data}. 
The blue line is the expected time over threshold distribution for a signal from the Lead-Glass block
without noise and the red histogram is the result with the 300 $\mu$V noise. 
The sinusoidal noise induces a random shift on the measured leading edge or 
trailing edge alone but correlates
them between each other - if the leading edge is crossed predominantly when the 
phase of the noise is $\pi/2$ the trailing edge is crossed when the phase is close to $3\pi/2$. 

\begin{figure}[htb]
\centering
\includegraphics[width=16cm]{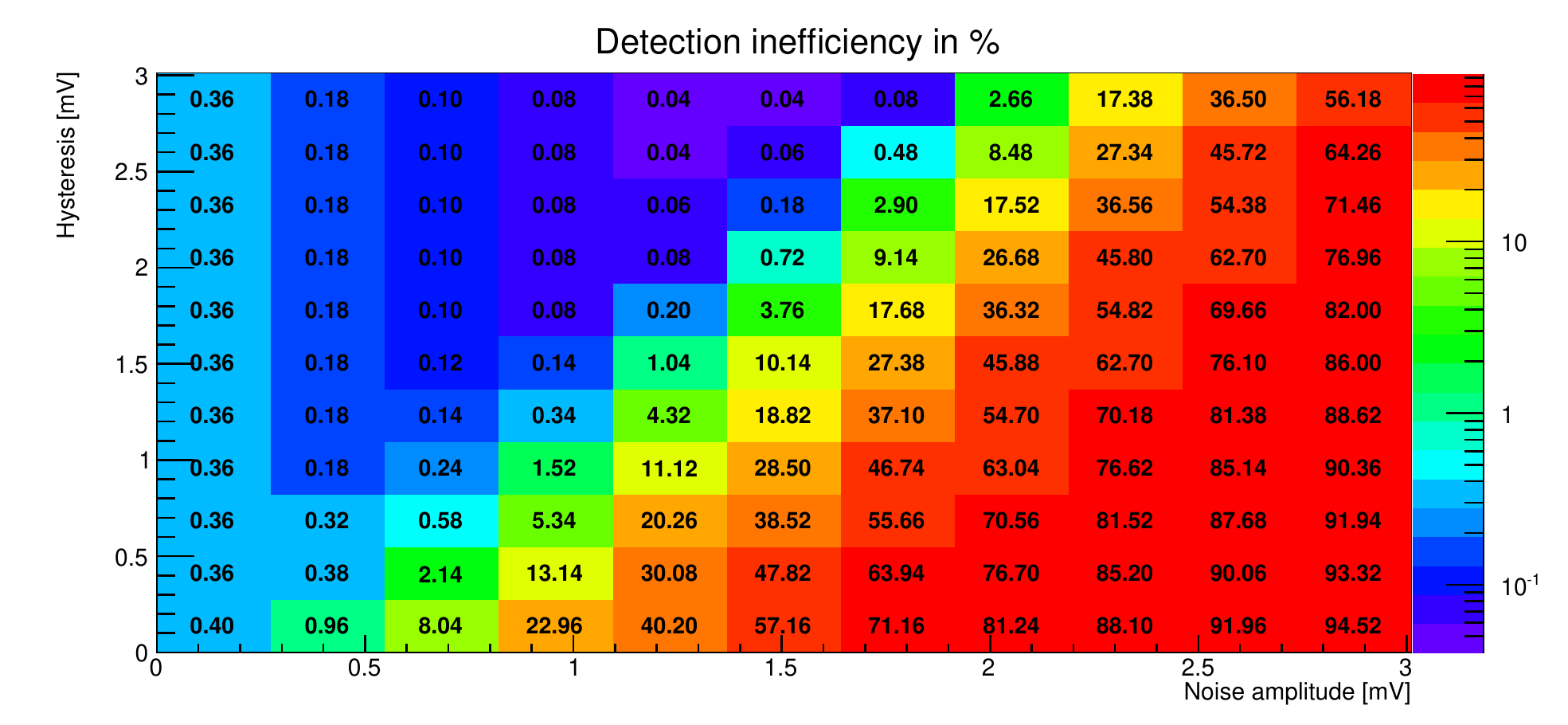}
\caption{Dependence of the inefficiency on the value of the hysteresis as function of the amplitude of the
input noise. An efficient hit is defined if the first leading and trailing edge form time over threshold longer than 5 ns.}
\label{fig:rel-noise-hyst}
\end{figure}

In a general setup the inclusion of noise doesn't change the detection efficiency for a signal. 
In the case where there is a minimal detectable (5 ns in the case of  HPTDC)
the noise can induce inefficiency through artificially decreasing the ToT of the signals 
that are close to the threshold.
If the period of the noise is smaller than the dead time of the TDC the situation becomes critical as
in the case of HPTDC with  300 MHz noise. 
The inefficiency is partially recovered by the presence of hysteresis shifting forward the trailing edge
but the effect could be dramatic as shown in Fig. \ref{fig:rel-noise-hyst}. 

The present studies underline the importance of the hysteresis 
to diminish the dependence on the external noise. 
The minimal hysteresis one should consider when employing ToT based solutions
for readout electronics is 50\% higher than the expected sinusoidal noise. 

The noise induces an additional uncertainty to the measurement of the 
time over threshold which combines with the uncertainty from the underlying physics
process - the shower development and primary charge generation. 
While the uncertainty of the latter usually scales as the square root of the 
deposited energy (due to linear scale of the primary charges), the uncertainty due to the noise
is constant in the time over threshold variable - $\Delta \Theta = const$. 
$\Delta \Theta$ can be assumed to be half of the distance between two consecutive clusterisation peaks,
as seen in Fig. \ref{fig:sim-real-data}. 
Then the physics quantity, the energy ($E\sim A$), will acquire an additional 
constant term in the resolution dependence as a function of energy
\begin{equation}
 \frac{\Delta E}{E} (E)_{ToT} = \frac{\Delta A}{A} (E)_{ToT} = b*\Delta \Theta = const.
\end{equation}

This term could be as high as tens of percent ($ b \sim 0.1$ ns$^{-1}$ 
and $\Delta \Theta\sim 1.5$ ns in this study)
 and will be the dominant one, especially in the measurement of the 
electromagnetic showers where the stochastic and the noise terms are usually quite low.

\section{Conclusions}

A comb-like structure was identified in the time over threshold
distribution in cosmic ray data for a first time, 
was explained to be caused by the pick up of high frequency low amplitude noise,
and was modeled successfully. The effect should be taken into account
by every detector readout system aiming to use ToT as a measurable quantity to 
describe the data from the detector. In the case of the Large Angle Vetoes readout
system additional precautions were taken (better cable shielding, extra noise 
filtering in the crate power supply) to decrease the level of the noise
to an acceptable level, which does not degrade the efficiency 
of the system.

\section*{Acknowledgments}
The present work was performed at the Laboratori Nazionali di Frascati, INFN. The authors are 
indebted to Antonella Antonelli, Matthew Moulson and Tommaso Spadaro for the 
pleasure of the joint work and the valuable discussions on the data and its interpretation. 
The time over threshold board was developed by Gianni Corradi and the authors would 
like to thank him for the useful discussions.

\end{document}